# Detection of excited state absorption cross-section of porphyrin through cw and femto-second laser pump-probe technique

A. Srinivasa Rao[1*], Alok Sharan[1], N Venkatramaiah[2], and R Venkatesan[2]

[1]Department of Physics, Pondicherry University, Puducherry, 605014, India
[2]Department of Chemistry, Pondicherry University, Puducherry, 605014, India
[3]Photonic Sciences Lab, Physical Research Laboratory, Ahmedabad 380009, Gujarat, India
*Email: asvrao@prl.res.in, sri.jsp7@gmail.com

**Abstract**
We report on direct detection of excited states absorption cross-section using dual wavelength pump-probe technique. Also, we experimentally demonstrate using porphyrin composite molecules (porphyrin derivatives such as 5,10,15,20 - meso-tetrakis phenyl porphyrin ($H_2TPP$), 5,10,15,20 - meso-tetrakis(4-hydroxyphenyl) porphyrin ($H_2TPP(OH)_4$)). The cw laser at 761 nm wavelength is used as a pump to maintain excited state population. Changes in the population of excited states lead to the change in transmission are monitored using femto-second probe pulses of 130 fs width and repeated at a 1kHz rate with central wavelength around 800 nm. Transmittance changes due to excited state population are modeled using rate equation approach. The effect of the absorption on the transmitted pulse shape has been discussed as a function of fluence. Obtained excited state (triplet transitions involving $T_1 \rightarrow T_2$ energy levels) absorption cross-sections of $H_2TPP$ and $H_2TPP(OH)_4$ doped boric acid glass (BAG) films are $4.9 \times 10^{-18}$ cm$^2$ and $1.2 \times 10^{-17}$ cm$^2$ respectively.

## 1. Introduction

Nonlinear absorption plays a major role in the nonlinear material applications. The nonlinear nature of absorption in the materials take place in two ways: either the absorption increases [1-3] or decreases [4-7] with increasing the excitation intensity. Such an anomalous nonlinear behavior in the absorption can be used for pulse shaping and can prevent the damage of optical devices from the sudden changes in the high power laser intensities [8-10]. Further, it is well established that nonlinear absorption has been employed to enhance the determination optical spectroscopic parameters of materials [11-13]. However, the nonlinear Spectroscopic technique involving saturation of levels and measuring the excited state dynamics of molecules using pump-probe technique is used to obtain information of the levels hitherto inaccessible by the conventional techniques [14-17]. While the pump beam injects the population of the molecules in its desired excited state, the probe beam is to induce and monitor the changes in the population involving excited triplet state energy levels. The pump-probe technique helps us to understand the excited state dynamics without any effect on the material optical properties [18-20]. The advantage of pump-probe over other spectroscopic techniques is that we can measure the non-degenerate wavelength spectroscopic parameters through non-degenerate pump-probe technique [21], where pump and probe derived from two different laser sources. The wide range of flexibility of pump-probe technique in terms of wavelength and pulse width of the laser beams allow them to use in wide range of electromagnetic spectrum in the process of materials characterization [22-26].

Here we report, the non-degenerate pump-probe technique with cw (continuous wave) laser at 671 nm as a pump and 130 fs (femto-second) pulses@1kHz repetition rate at λ=800nm central wavelength as probe. To experimentally demonstrate our method, we have use porphyrin composite molecules [27,28] (porphyrin derivatives such as 5,10,15,20 - meso-tetrakis phenyl porphyrin ($H_2TPP$), 5,10,15,20 - meso-tetrakis(4-hydroxyphenyl) porphyrin ($H_2TPP(OH)_4$)) doped in boric acid glass (BAG) as sample and estimated the excited state absorption cross-section. The porphyrin dye molecules doped in BAG sandwiched films were prepared by rapid quenching of the melt technique. The full details of the sample preparation, and linear as well as nonlinear characterizations can be found elsewhere [29]. The thickness of $H_2TPP$ and $H_2TPP(OH)_4$ doped in boric acid glass films are 88 μm and 67 μm respectively.

## 2. Theory

Energy level diagram of organic dyes like porphyrin molecules is represented by Jablonskii diagram, where the levels involving spin multiplicity viz singlet or triplet are as shown in figure 1. The number density of molecules in each state "i" is considered to be $N_i$ ($N_0$, $N_1$, $N_2$ and $N_3$ are respective populations in the $S_0$, $T_1$, $S_1$, and $T_2$ states). The total number density of molecules is given as $N=\Sigma N_i$ and the fractional number density of each state is given by $n_i=N_i/N$, with $\Sigma n_i=1$. The pumping rate from $i^{th}$ to $j^{th}$ level is $W_{ij}=(I/h\nu)\sigma_{ij}$ and the decay time from $j^{th}$ to $i^{th}$ level is $\tau_{ji}$. Molecules are pumped from ground state $S_0$ to triplet state $T_1$ via $S_1$ by 671 nm.

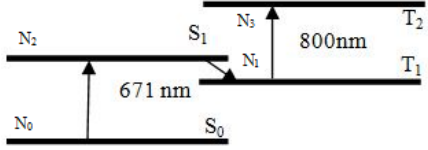

Figure 1. Energy level diagram of H$_2$TPP doped BAG for pump-probe experiment.

To investigate the population in each energy level, we have adopted the rate equation aproach [30, 31]. Using rate equations, we can efficiently understand the population transfer among the energy levels participated in the absorption process. In the presence of cw laser excitation, depending up on the excitation intensity, the population transfer takes place among the energy levels S$_0$, S$_1$, T$_1$, and T$_2$ energy levels. Thus, the rate equations populations in the energy levels are given by

$$\frac{dn_0}{dt} = -W_{02}(n_0 - n_2) + \frac{n_1}{\tau_{10}} + \frac{n_2}{\tau_{20}} \quad (1a)$$

$$\frac{dn_1}{dt} = -W_{13}(n_1 - n_3) - \frac{n_1}{\tau_{10}} + \frac{n_2}{\tau_{21}} + \frac{n_3}{\tau_{31}} \quad (1b)$$

$$\frac{dn_2}{dt} = W_{02}(n_0 - n_2) - \frac{n_2}{\tau_{20}} - \frac{n_2}{\tau_{21}} \quad (1c)$$

$$\frac{dn_3}{dt} = W_{13}(n_1 - n_3) - \frac{n_3}{\tau_{31}} \quad (1d)$$

Here, our cw pump source is pumping the optical energy continuously. Hence, the population comes to the study state and it leads to the $dn_i/dt=0$. As a consequence of this, equations 1 is independent of time and analytically solved to obtain population in each state ($n_i$) as a function of intensity. We use I$_{CW}$=6×10$^4$ mW/cm$^2$ intensity to pump the molecules to the triplet state T$_1$ without accumulating any population in the triplet state T$_2$ [32]. For comparative study, we have doped the same concentration (4×10$^{-5}$ M) of H$_2$TPP and H$_2$TPP(OH)$_4$ in BAG thin films. The fractional population that was pumped to triplet level (T$_1$) for studying excited state absorption cross-section is obtained from rate equations 1 and used as absorptive species for femto second laser excitation.

The population redistribution between the triplet states (T$_1$, and T$_2$) in the presence of 130 femto-second laser pulses was studied to obtain triplet to triplet (T$_1$→T$_2$) absorption cross-section. The redistribution of molecules among the triplet states is studied as a function of pulse fluence by varying the average power of femto-second laser beam. The redistributed population between the triplet levels under femto-second laser excitation can be obtained through rate equations, which are given as

$$\frac{dn_1}{dt} = -\frac{\sigma_{13}I(t)}{h\nu}(n_1 - n_3) \quad (2a)$$

$$\frac{dn_3}{dt} = \frac{\sigma_{13}I(t)}{h\nu}(n_1 - n_3) \quad (2b)$$

The intensity distribution within the femto-second laser pulse is given by

$$I(t) = I_p \exp(-2r^2/\omega^2)\exp(-4\ln 2 t^2/\tau_{FWHM}^2) \quad (3)$$

The fractional populations in the T$_1$ and T$_2$ states are obtained by integrating over pulse-widths

$$n_1 = \frac{1}{2}\left[1 + \exp\left(-2F_S^{-1}\int_{-\infty}^{+\infty}I(t)dt\right)\right] \quad (4a)$$

$$n_3 = \frac{1}{2}\left[1 - \exp\left(-2F_S^{-1}\int_{-\infty}^{+\infty}I(t)dt\right)\right] \quad (4b)$$

Here, F$_S$=h$\nu$/$\sigma_{13}$ is the saturation fluence. The absorption coefficient in terms of absorption cross-section is $\alpha$=N(n$_1$-n$_3$)$\sigma_{13}$. The incident pulse energy on the material is given by integrating the intensity in the cross-sectional area and pulse-width.

$$E(0) = I_p \frac{\pi\omega^2}{2}\sqrt{\frac{\pi}{4\ln 2}}\tau_{FWHM} \quad (5a)$$

The transmitted energy through the sample of thickness L is given by

$$E(L) = I_p \frac{\pi\omega^2}{2}\sqrt{\frac{\pi}{4\ln 2}}\tau_{FWHM}\exp(-\alpha L) \quad (6b)$$

## 3. Experiment

The layout of experimental setup used for pump-probe is illustrated in figure 2. Pump source, cw laser of Model SDL-671-120T emitting at a 671 nm wavelength with average power 150 mW directed towards dichroic mirror DM$_1$ through mirror M$_3$. The probe beam is derived from TOPAS, which is pumped by Ti:sapphire femto-second laser. The femto-second laser pulses used in the experiment have the pulse width of 130 fs at 800 nm central wavelength with a repetition rate of 1kHz. The probe beam is diverted towards dichroic mirror DM$_1$ using mirrors M$_1$, and M$_2$. Wavelength filters (WF) are used to remove other residual wavelengths generated in the TOPAS. Iris apertures A$_1$, and A$_3$ are used to prevent the laser from back reflections of optical components. The femto-second and cw laser beams are collinearly incidents on the sample by use of a dichroic mirror (DM$_1$). Lenses L$_1$ and L$_2$ are used to focus the respective pump and probe beams on the sample. Another dichroic mirror (DM$_2$) is inserted before detector (PD$_3$) to remove the cw laser beam. We took care to avoid spurious reflections from the optical elements by inserting irises at appropriate positions in the experimental setup. The transmitted data is collected using detector D$_3$ and monitored with the reference photo-diodes (PD$_1$ and PD$_2$). We have varied the incident intensity of probe using neutral density (ND) filters and studied the transmitted intensity as a function of incident intensity.

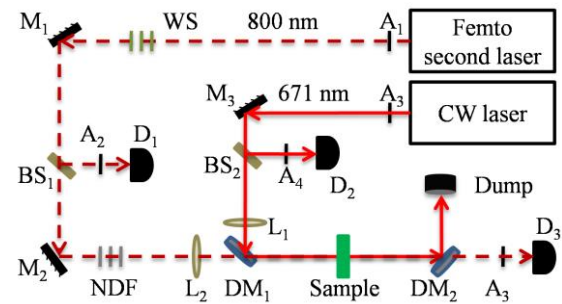

Figure 2. Schematic layout of pump-probe experimental setup: $M_i \rightarrow$ Mirror, $DM_i \rightarrow$ Dichroic Mirror, WF $\rightarrow$ Wave length filters, $F_i \rightarrow$ Neutral density filters, $PD_i \rightarrow$ Photo diode, $A_i \rightarrow$ Iris, $BS_i \rightarrow$ Beam splitter.

The variation in the transmittance of femto-second laser beam through both $H_2TPP$ and $H_2TPP(OH)_4$ as a function of incident fluence are presented in figure 3. The generated experimental data were fitted with femto-second laser transmittance curves obtained from equations 4, and 5. From the theoretical fits of experimental data, we obtain the excited state absorption cross-section ($\sigma_{13}$) for $H_2TPP$ and $H_2TPP(OH)_4$ doped BAG as $4.9\times10^{-18}$ cm$^2$ and $1.2\times10^{-17}$ cm$^2$ respectively. The -(OH) group addition to the $H_2TPP$ molecule increased the excited state absorption cross-section. As a result of this, the transmittance at the linear range for $H_2TPP(OH)_4$ is less than the $H_2TPP$. The large absorption cross-section contained molecules get saturate at low intensity as compared with low absorption cross-section molecules since the stimulated absorption directly proportional to the absorption cross-section. The saturation in terms of fluence is inversely proportional to absorption cross-section ($F_S = h\nu/\sigma_{13}$). One can observe this behavior in figure 3, The $H_2TPP(OH)_4$ curve saturated at low fluences due to its large absorption cross-section as compare with $H_2TPP$.

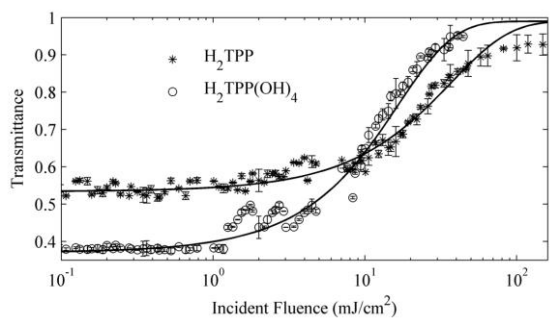

Figure 3. Transmittance plot of $H_2TPP$ and $H_2TPP(OH)_4$ doped BAG due to excited state (triplet $T_1$ to $T_2$) absorption: scatters are experimental data and lines are theoretical data.

The effect of absorption on the transmitted pulses as a function of fluence is depicted in figure 4. To qualitatively visualize the changes in the pulse shape, we comparatively plotted the transmitted curves with respect to incident pulse (red curve). In both the cases of $H_2TPP$ and $H_2TPP(OH)_4$ molecules, the shape of the transmitted pulses from the sample are changed. In all the cases, one can see that the leading edge of the pulse absorbed more as compared with the trailing edge. Initially, the population present in the first triplet state ($n_1 - n_3 = n_1$) and the absorption coefficient is given by $\alpha = Nn_1\sigma_{13}$. At trailing edge due to finite life time of excited states, the population in the second triplet state have the non zero value ($n_3 \neq 0$). Therefore, the absorption coefficient at trailing edge is $\alpha = N(n_1 - n_3)\sigma_{13}$. As a result of this scenario, The transmittance value at leading is less than the trailing edge. Effect of relative absorption cross-sections of molecules can be seen in the pulse transmittance. At 5 mJ/cm$^2$ fluence, the transmitted pulse from $H_2TPP(OH)_4$ distorted in a greater amount than the pulse from $H_2TPP$ due to large absorption cross-section of $H_2TPP(OH)_4$ (Solid lines in figure 4). As like in the transmittance (Fig. 3), at 30 mJ/cm$^2$ fluence, the peak intensity of the transmitted pulse for $H_2TPP(OH)_4$ is saturated faster than the $H_2TPP$ due to the condition: saturation fluence of $H_2TPP(OH)_4$ < saturation fluence of $H_2TPP$. At complete saturation, the transmittance curve comes to the symmetric intensity distribution as like incident pulse.

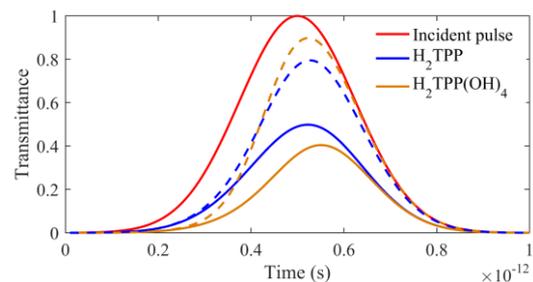

Figure 4. Transmittance of the femto-scond Pulse at different fluence excitation of $H_2TPP$ and $H_2TPP(OH)_4$ doped BAG: straight line corresponds to 5 mJ/cm$^2$ and dashed line corresponds to 30 mJ/cm$^2$.

**4. Conclusion**
We have discussed the pump-probe technique using cw laser as pump and femto-second laser as a probe to estimate the excited state absorption cross-section. The rate equation technique is used as a theoretical model to understand and analyze the pump and probe beams on the material in terms of populations in the energy levels. We have used $H_2TPP$ and $TPP(OH)_4$ molecules with the same concentration and prepared under similar experimental conditions in the pump-probe transmittance. Extracted the excited state absorption cross-sections of these molecules by analyzing the probe transmittance curve by rate equations. The advantage of this kind of experimental technique is that there will not be any effect of probe beam on the pumping rate because the final state of pumping and the initial state of probing are different. Using this technique, we can study the excited states spectroscopic parameters, where energy levels are not equally separated (resonant at different wavelengths).

**Acknowledgement**: Authors thanks CIF Pondicherry University for providing femto-second laser facility.